\documentstyle[epsfig,sprocl]{article}
\def\Journal#1#2#3#4{{#1} {\bf #2}, #3 (#4)}

\def\PLB{{\em Phys. Lett.}  B}

\def\ra{\rightarrow}
\def\be{\begin{equation}}  
\def\ee{\end{equation}}
\def\bea{\begin{eqnarray}}
\def\eea{\end{eqnarray}}
\newcommand{\lsnd}{LSND } 
\newcommand{\nue}{neutrino } 
\newcommand{\nues}{neutrinos }

\newcommand{\noszp}{neutrino oscillations. }
\newcommand{\osz}{oscillation }
\newcommand{\oszsp}{oscillations. }
\newcommand{\oszs}{oscillations }

\newcommand{\sk}{Superkamiokande }

\newcommand{\delm}{\mbox{$\Delta m^2$} }

\newcommand{\nel}{\mbox{$\nu_e$} }
\newcommand{\bnel}{\mbox{$ \bar \nu_e$} }
\newcommand{\bnmu}{\mbox{$ \bar \nu_{\mu}$} }
\newcommand{\nmu}{\mbox{$\nu_\mu$} }
\newcommand{\ntau}{\mbox{$\nu_\tau$} }
\newcommand{\sint}{\mbox{$sin^2 2\theta$} }
\newcommand{\lbls}{long baseline experiments }

\begin{document}
\title{STATUS AND PERSPECTIVES OF NEUTRINO OSCILLATION SEARCHES
\footnote{to appear in Proc. 
{\it 6$^{th}$ Int. Symposium on
Particles, Strings and Cosmology (PASCOS'98)}, Boston, March 1998}}
\author{ K. ZUBER}
\address{Lehrstuhl f\"ur Experimentelle Physik IV, Universit\"at Dortmund,\\
Otto-Hahn Str. 4, 44287 Dortmund, Germany}
\maketitle
\abstracts{The current status of \nue \osz searches with reactors and accelerators is reviewed.
An outlook, especially
on future long baseline \nue \osz projects, is given.}
\section{Introduction}
The existence of massive \nues opens up a variety of new phenomena which could
be investigated by experiments. One of these is \nue \oszsp 
In the simplified picture of two flavour \oszs they can be parametrized by two parameters, \sint and $\Delta
m^2$. 
While $\sint$ describes the amplitude of the
oscillation, $\delm = m_2^2 -m_1^2$ determines the oscillation length $L$ given in
practical units as
\be
L = \frac{4\pi E \hbar}{\Delta m^2 c^3} =
2.48 (\frac{E}{MeV})(\frac{eV^2}{\Delta m^2}) \quad m
\ee
As can be seen, \oszs do not allow an absolute mass measurement and \nues must not be exactly degenerated. 
For a general discussion of direct mass bounds and on physics with massive \nues see
\cite{phrep}. 
From first principles, there is no preferred region
in the $\delm - \sint$
parameter space and therefore the whole has to be investigated experimentally.\\
On earth, two artificial \nue sources exist in form of nuclear power reactors and accelerators.
For a more detailed overview see \cite{hd}.
\section{Reactor experiments}
\begin{figure}[hhh] 
\begin{center}
\begin{tabular}{ll}
\epsfig{file=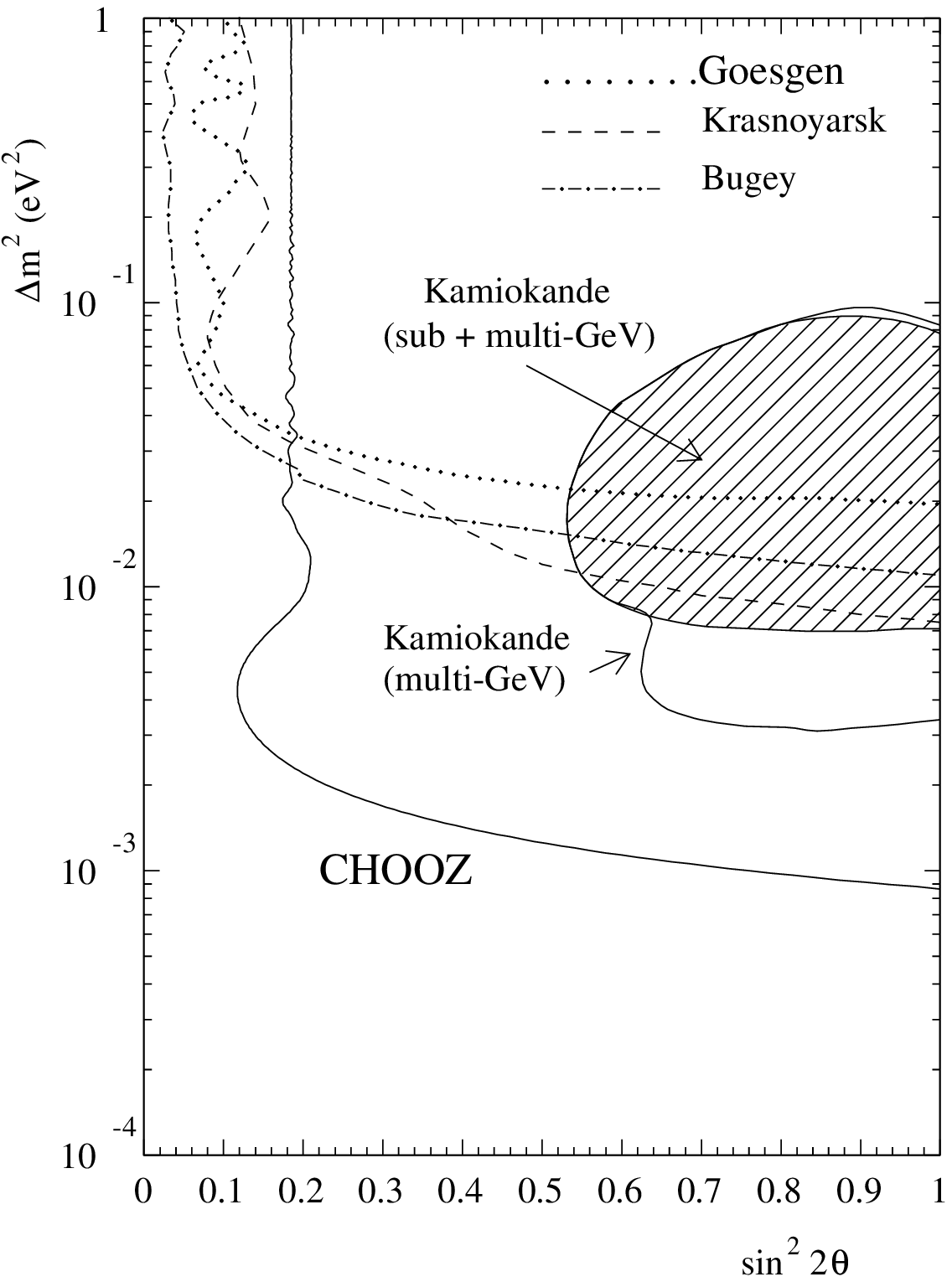,width=5.8cm,height=5.8cm} &
\epsfig{file=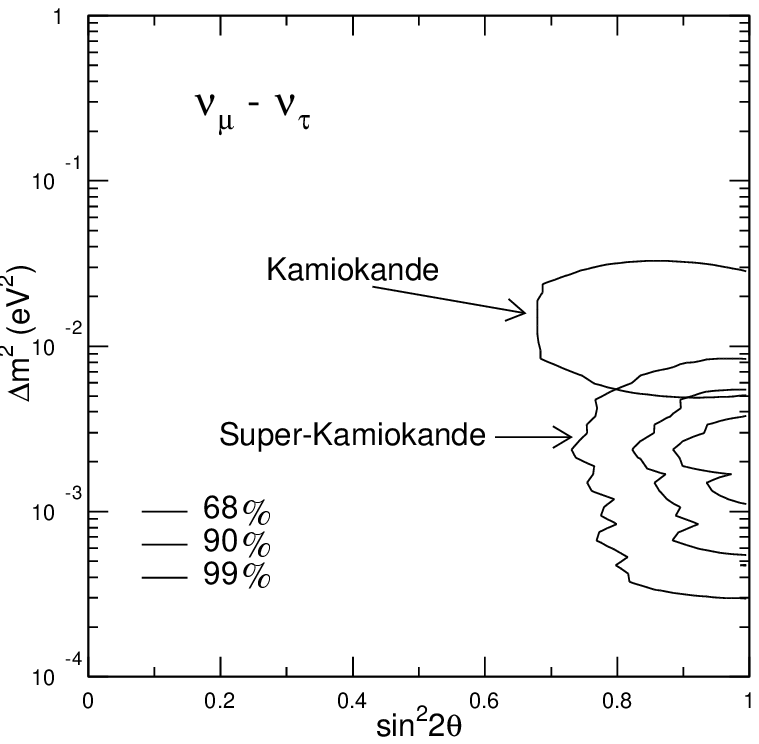,width=5.5cm,height=5.5cm} 
\end{tabular}
\end{center}
\caption{\label{chores} \it Left: Exclusion plot for \bnel - $\bar{\nu}_X$ \osz as given
by the
CHOOZ-results and other reactor experiments. Also shown are the allowed
regions from
atmospheric \nues given by Kamiokande. As can be seen, the complete
region is excluded. 
Right: 68 \%, 90 \% and 99 \% confidence intervals for \nmu - \ntau \osz necessary to
explain
the atmospheric \nue deficit as
observed by Superkamiokande in their 33.0 kt$\cdot$y data sample. The Kamiokande
90 \% region is shown for comparison.}
\end{figure}

 Reactor experiments are looking for \bnel $\ra \bar {\nu}_X$ disappearance.  Reactors are a
source of MeV \bnel, due to the fission of nuclear fuel.  Experiments typically try to measure
the positron spectrum which can be deduced from the \bnel - spectrum and either compare it
directly to the theoretical predictions or measure it at several distances from the reactor
and search for spectral distortions. Both types of experiments were done in the past. The
detection relies on the reaction 
\be 
\label{gl1} \bnel + p \ra e^+ + n 
\ee 
with an energy
threshold of 1.804 MeV.  Different strategies are used for the detection of the positron and
the neutron. Normally, coincidence techniques are used between the annihilation photons and
the neutrons which diffuse and thermalise within 10-100 $\mu$s and materials like Gd are then
used for neutron-capture.  The most recent experiment is CHOOZ in France \cite{chooz}. 
Compared to previous experiments, this detector has some advantages.  First of all, the
detector is located underground with a shielding of 300 mwe, reducing the background due to
cosmics by a factor of 300. Moreover, the detector is about 1030 m away from the reactor (more
than a factor 4 in comparison to previous experiments)  enlarging the sensitivity to smaller
$\Delta m^2$. In addition, the main target has about 4.8 t of a specially developed Gd-loaded
scintillator and is therefore much larger than those used before. First results can be seen in
Fig.\ref{chores}.\\ 
An experiment with similar goals is the Palo Verde (former San Onofre)
experiment \cite{pave} near Phoenix, AZ (USA). It consists of a 12 t liquid scintillator also
loaded with Gd. The experiment is located under a shielding of 46 mwe in a distance of about
750 (820) m to the reactors. The experiment started data taking recently.\\
A first long-baseline reactor
experiment (KamLAND) \cite{kamland} using a 1000 t liquid scintillator detector at the Kamioka
site in a distance of 150 km to a reactor is approved by the Japanese Government. It could
start data taking in 2000.  
\section{Accelerators} 
Accelerators typically produce \nue beams
by shooting a proton beam on a fixed target.  The produced secondary pions and kaons decay and
create a \nue beam dominantly consisting of $\nu_{\mu}$.  The detection relies on charged
current reactions $\nu_i N \ra i + X \quad (i= e, \mu, \tau$), where N is a nucleon and X the
hadronic final state.  Depending on the intended goal, the search for \oszs therefore requires
a detector which is capable of detecting electrons, muons and $\tau$ - leptons in the final
state.  Accelerator experiments are mostly of appearance type working in the channels \nmu -
$\nu_X$ and \nel - $\nu_X$.  
\subsection{Accelerators at medium energy} 
\begin{figure}[hhh] 
\begin{center} 
\epsfig{file=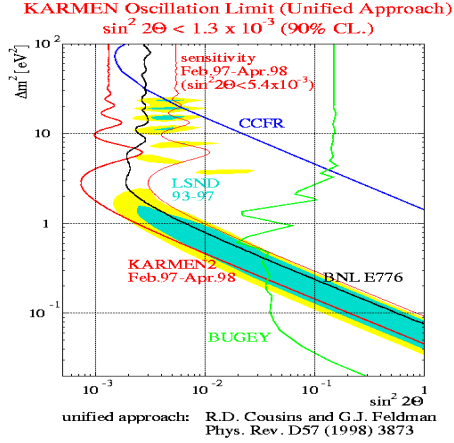,width=6cm,height=6cm}
\end{center} 
\caption{\label{lsndev} \it Region of evidence for \bnmu - \bnel \oszs from \lsnd
together with already excluded parts from KARMEN, E776, CCFR and the Bugey reactor
experiment.}
\end{figure}

At present there are
two experiments running with \nues at medium energies ($E_\nu \approx $ 30 - 50 MeV) namely
KARMEN \cite{karmen} and \lsnd \cite{lsnd}. \lsnd finds evidence for \oszs in the 
\nel - \nmu channel for pion
decay at rest and in flight.  
To improve the sensitivity for \osz searches by reducing the neutron background KARMEN
constructed
a veto shield against atmospheric muons which has been in operation since Feb.1997 and is
surrounding the whole detector. 
The limits reached so far and the \lsnd evidence
are shown in Fig.\ref{lsndev}. The new analysis of KARMEN seems to be
in contradiction with the LSND evidence. While \lsnd will stop data aquisition after 1998,
KARMEN will continue until 2000.  
\newline
To test the \lsnd region of evidence several new projects are planned.  The Fermilab 8 GeV
proton booster offers the chance for a \nue experiment as well which could start data taking
in 2001. It would use part of the \lsnd equipment and will consist of 600 t mineral oil
to be located 500 m away from the \nue source (MiniBooNE) \cite{boone}.  An extension
using a
second detector at 1000m is possible (BooNE).  An increase in sensitivity in the \nel - \nmu
\osz channel could also be reached by a proposed experiment at the CERN PS \cite{cernps} or if
there is a possibility for \nue physics at the planned European Spallation Source (ESS)  or
the National Spallation Neutron Source (NSNS) at Oak Ridge which might have a 1 GeV proton
beam around 2004.  
\subsection{Accelerators at high energy} 
High energy accelerators provide \nue beams with an average energy in the GeV
region.  Here, at present especially CHORUS and NOMAD at CERN are providing new limits
\cite{chorus}.  Both experiments are 823 m (CHORUS) and 835 m (NOMAD) away from a beam dump
and designed to improve the existing limits on \nmu - \ntau \oszs by an order of magnitude.
The present limits (Fig.\ref{sum}) for large \delm are \cite{chorus} 
\bea 
\sint < 1.3 \times 10^{-3} \quad (90 \% CL) \quad (CHORUS)\\ 
\sint < 2.2 \times 10^{-3} \quad (90 \% CL) \quad (NOMAD)  
\eea 
The final goal is to reach a sensitivity down to \sint $\approx 2 \times
10^{-4}$ for large $\Delta m^2$.  Having a good electron identification NOMAD also offers the
possibility to search in the \nel - \nmu channel.  
While the CHORUS data taking is finished, NOMAD continues 1998.  
\section{Future accelerator experiments} 
\begin{figure}[hhh]
\begin{center}
\epsfig{file=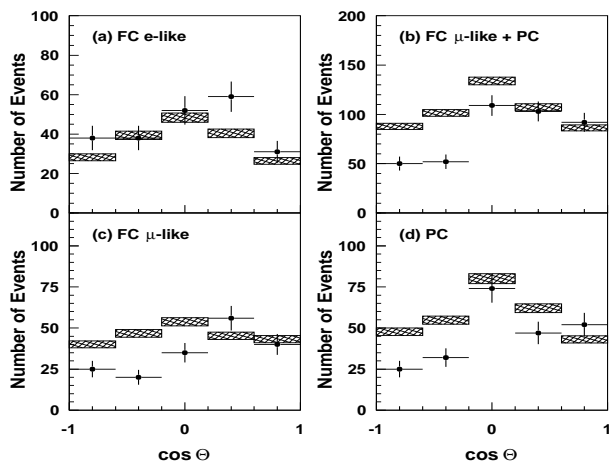,width=8cm,height=6cm}
\end{center}
\caption{\label{superk} \it Zenith angle distribution of a) FC e-like events b) FC
$\mu$-like and PC events c) FC $\mu$-like and d) PC events as observed with \sk . 
The statistical significance corresponds to 25.5 kt$\cdot$y.
Vertical downward going events correspond to $cos \theta=1$, vertical upward going to
$cos \theta=-1$. The shaded boxes are the MC predictions including the statistical
uncertainties.}
\end{figure}

Possible future ideas split into two groups depending
on the physical
goal. One part is focussing on improving the existing bounds in the eV-region by another order
of magnitude with respect to CHORUS and NOMAD and to investigate the \lsnd evidence.  Other
groups plan to increase the source - detector distance to probe smaller \delm and to be
directly comparable to atmospheric scales. The last point is of special importance
because of the recent claim of evidence for \nmu - \ntau \oszs in the atmospheric \nue
data by Super-Kamiokande
\cite{sk}, which are shown in Fig.\ref{chores} and \ref{superk}.  
\subsection{Short and medium baseline experiments}
\begin{figure}[hhh]
\begin{center} 
\epsfig{file=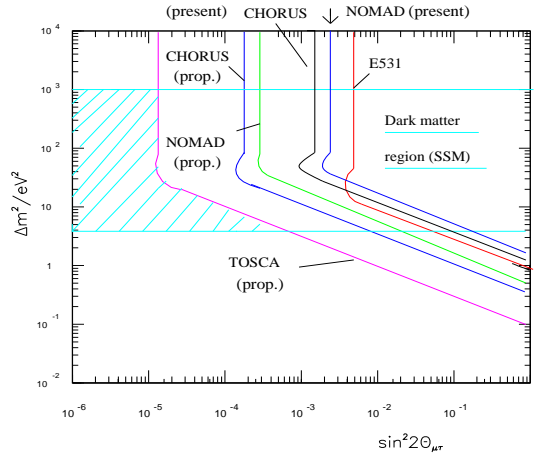,width=7cm,height=6cm}
\end{center} 
\caption{\label{sum} \it \nmu - \ntau exclusion plot showing the present limits
of CHORUS and NOMAD as well as the proposed limits of CHORUS, NOMAD and TOSCA. The shaded
area
corresponds to the signal region if there is a \ntau in the eV-range as motivated by
dark matter considerations
and assuming the quadratic see-saw-mechanism (SSM).} 
\end{figure}

Ideas exist for a next generation of short or medium baseline experiments. At CERN the
follow up
is TOSCA \cite{tosca}, combining features of NOMAD and CHORUS. The idea is to use 2.4
tons of emulsions together with large silicon microstrip detectors within the NOMAD magnet. 
For TOSCA the option to extract a \nue beam at lower proton energies (350 GeV) at the CERN SPS
exist.  
The proposed sensitivity in the \nmu - \ntau channel is around 2$\times 10^{-5}$ for large
\delm (\delm $>100 eV^2$) (Fig.\ref{sum})
and data taking could start at the beginning
of the next century.  Also proposals for a medium baseline search exist
\cite{medium,icarus}. The
present CERN \nue beam is coming up to the surface again in a distance of about 17 km away
from the beam dump offering the chance for an experiment there.  
\subsection{Long baseline experiments} 
Several accelerators and underground laboratories around the world offer the
possibility to perform \lbls, a search which is strongly motivated by the new results on
atmospheric \nues . Because the region of evidence moved towards smaller \delm by roughly one
order of magnitude
most of the experiments reconsider their design.
\smallskip\\ 
{\it KEK - \sk}:  The first of these experiments
will be the KEK-E362 (K2K) experiment \cite{keksk} in Japan sending a \nue beam from KEK to
\sk. The distance is about 235 km. A near detector, about a 1 km away from the beam dump, will
consist
of two detectors, a 1 kt water Cerenkov-detector and a further detector consisting of a
SciFi/water target followed by trigger counters, a lead glass calorimeter and a muon-detector.
They
will serve as a reference and measure the \nue spectrum. The \nue beam with an average energy of
1.4
GeV is produced by a 12 GeV proton beam dump. The detection method within \sk will be
identical to that of their atmospheric \nue detection.  The beamline should be finished by the
end of 1998 so the experiment could start data taking in 1999.  The experiment is of
disappearance type. However an upgrade of KEK to a 50 GeV proton beam is planned, which could
start producing data around 2004 and would allow \ntau-appearance searches.  
\smallskip\\ 
{\it Fermilab - Soudan}:  A \nue program is also associated with the new Main Injector at
Fermilab. The long baseline project will send a \nue beam to the Soudan mine about 735 km away
from Fermilab. Here the MINOS experiment \cite{minos} will be installed. It consists of a
near detector located at Fermilab and a far detector at Soudan. The far
detector will be made of 8 kt magnetized Fe toroids in 600 layers with 2.54 cm thickness
interrupted by about 32000 m$^2$ active detector planes in form of plastic scintillator
strips 
to get the
necessary tracking informations. An additional hybrid emulsion detector for
$\tau$-appearance is also under consideration. The final beam line layout is still under
investigation. The project could start at the beginning of next century.  
\smallskip\\ 
{\it CERN - Gran Sasso}:  A further program in
Europe considers \lbls using a \nue beam from CERN to Gran Sasso Laboratory.  The distance is
about
732 km. Several experiments have been proposed for the \osz search. The first proposal is the
ICARUS experiment \cite{icarus} which will be installed in Gran Sasso anyway for the search of
proton decay and solar neutrinos by using a liquid Ar TPC. A prototype of 600 t is approved
for installation in 1999. An upgrade to about 3 kt is foreseen.  A second proposal, the NOE
experiment \cite{noe}, plans to build a
giant combination of lead - scintillating fibre and transition radiation detectors  
with a total mass of 6.7 kt, followed by a module for muon identification. A third proposal is
a 125 kt water-RICH
detector (AQUA-RICH) \cite{tom}, which could be installed outside the Gran
Sasso tunnel. It could be independently used for measuring atmospheric neutrinos. Finally
there
exists a proposal for a 750 t iron-emulsion sandwich detector
(OPERA) \cite{niwa}. It would use thin iron plates as target as well as emulsion sheets for
tracking purposes. The $\tau$-decay would happen in an air gap between the emulsions. 
\section{Summary and Conclusion} 
Massive \nues allow a wide range of new phenomena in \nue
physics, especially that of \noszp Evidence for such \oszs comes from solar neutrinos,
atmospheric \nues and the \lsnd experiment. The \delm regions allowed are
around 1 eV$^2$ (\lsnd), around $10^{-2} -10^{-3} eV^2$ (atmospheric) and
$10^{-5} eV^2$ (MSW, solar) or $10^{-10} eV^2$ (vacuum, solar). Terrestrial \nue
experiments in form of nuclear
reactors and high energy accelerators already exclude large parts of the parameter space
because of non-observation of \osz effects. Because the region of the MSW-solution for solar
\nues are out of range for terrestrial experiments, current and to a large extend future \osz
experiments are motivated by the atmospheric \nue deficit, an eV-\nue as dark matter candidate
and a proof of the \lsnd results. 
Long-baseline experiments are necessary to explore the atmospheric region of evidence but
might fail to explore it completely.

\end{document}